\documentclass[traditabstract,a4,letter]{aa}  

\usepackage{epsfig}
\usepackage{graphicx}
\usepackage{epstopdf}
\usepackage{color}  
\usepackage{array}   
\usepackage{units}  
\usepackage[breaklinks,colorlinks, citecolor=blue]{hyperref}
\usepackage{natbib}  
\usepackage{multirow}   
\usepackage{amsmath,amssymb}  
\usepackage[english]{babel}
\usepackage[latin1]{inputenc}
\usepackage[T1]{fontenc}
\usepackage{longtable,lscape}
\usepackage{footnote}

\usepackage{txfonts}
\usepackage[toc,page]{appendix} 

\def\be{\begin{equation}}
\def\ee{\end{equation}}
\def\ba#1\ea{\begin{align}#1\end{align}}

\def\mr{\mathrm}

\begin{document}


\title{Cosmological constraints from the mass accretion rate I:\\ galaxy cluster number count evolution}
\author{G.Hurier\inst{1}}

\institute{
1 Centro de Estudios de F\'isica del Cosmos de Arag\'on (CEFCA),Plaza de San Juan, 1, planta 2, E-44001, Teruel, Spain
\\
\email{hurier.guillaume@gmail.com} 
}

\abstract{Galaxy cluster number count has been proven to be a powerful cosmological probe. However, cosmological constraints established with galaxy cluster number count are highly dependent on the calibration of the mass-observable relations.\\
Thanks to its nearly mass independence the specific mass accretion rate of galaxy clusters is nearly insensitive to the calibration of mass-observable relations.
The study of galaxy cluster number count evolution allows to probe the galaxy cluster mass accretion history in the context of an homogenous Universe. 
In this paper, we use relative abundance matching to infer the galaxy cluster mass accretion rate (MAR) for $z \in [0.0,0.6[$. 
Then, we use the MAR to set cosmological constraints. 
We found that this cosmological probe is sensitive to $\sigma_8 \Omega_{\rm m}^{-0.3} H_0^{-0.2}$ whereas the galaxy cluster count is sensitive to $\sigma_8 \Omega_{\rm m}^{0.3}$.
We used the second $Planck$ catalog of Sunyaev-Zel'dovich sources and we derive $\sigma_8 \Omega_{\rm m}^{-0.3} H_0^{-0.2} = 0.75 \pm 0.06$.
This results is consistent with cosmological constraints derived from galaxy clusters number counts, angular power spectrum, and cosmic microwave background analyses.\\
Therefore, the MAR is a key cosmological probe that can break the $\sigma_8$-$\Omega_{\rm m}$ degeneracy and that is not sensitive to the calibration of the mass-observable relations and does not requires a parametric form for the galaxy cluster mass-function.
}


\authorrunning{G.Hurier}
\titlerunning{Cosmological constraints from mass accretion rate I}

\maketitle
  
\section{Introduction}

Galaxy clusters are the largest gravitationally bound structures in the Universe.
Therefore, their statistic is a tailored probe of the large scale structure growth with cosmic time.\\
Numerical simulations \citep[see e.g.,][]{tin08,wat12} have proven that galaxy cluster number count tightly scales with cosmological parameters.\\

Galaxy clusters observation is allowed by several probes: over-density of galaxies \citep{wen12,roz14}, weak lensing produced on background galaxies \citep{hey12,erb13}, X-ray emission produced by the hot gas of electrons within galaxy clusters through Bremsstrahlung radiation \citep{boh01}, and the thermal Sunyaev-Zel'dovich (tSZ) effect on the cosmic microwave background produced by the same population of electrons \citep{sun72}.  \\
The establishment of cosmological constraints from galaxy cluster abundance or angular power spectrum is now a well-developed activity.\\
Tight cosmological constraints have been derived from galaxy cluster number count \citep{has13,rei13,planckszc,dah16}.
However, galaxy cluster number count constraining power is limited by the accuracy of the mass-observable scaling relations \citep{planckszc}.\\

Contrary to galaxy cluster number counts, the galaxy cluster mass accretion rate (MAR) is nearly proportional to the galaxy cluster mass \citep[see e.g.,][]{cor15}. Therefore, the specific mass accretion rate (sMAR = MAR/$M$) is nearly independent of the mass.\\
Consequently, the utilization of the MAR to set cosmological constraints is nearly independent of the calibration of mass-observable scaling relations. However, it remains dependent of the slope of the mass-observable relations.
Additionally, the MAR increases with the matter density, $\Omega_{\rm m}$, but decreases with the amplitude of the matter fluctuations, $\sigma_8$. Whereas, galaxy cluster total number count increases with these two parameters.\\
The main limitation in using the galaxy cluster MAR comes from its measurement \citep{deb16}. Recent detections of the galaxy cluster splashback radius \citep{shi18} and virial shock \citep{hur19} are providing the opportunity to measure the galaxy cluster MAR \citep{shi16}.\\ 
However, the splashback radius is detected on stacked data, making its cosmological interpretation difficult, and the virial shock has been detected for one single cluster so far.\\

In this work, we use relative abundance matching on galaxy clusters detected via the tSZ effect to measure the average MAR of galaxy clusters from z = 0.6 to z = 0.0. 
This approach allows to isolate the cosmological information contained in the evolution of the galaxy cluster mass function.\\
Isolating such information is crucial to test the overall consistency of our understanding of large scale structure evolution and of the standard model of cosmology.\\
In particular, the abundance matching approach in the high-mass end of the mass function allows to have little sensitivity to the considered galaxy cluster sample selection function.\\
This probe is therefore particularly interesting for combination with cosmological constraints derived from galaxy cluster statistics that are not sensitive to selection effects such as: (i) the tSZ angular power spectrum \citep{rei12,sie13,geo15,planckszs}, (ii) the tSZ skewness \citep{wil12}, or (iii) the tSZ bispectrum \citep{Crawford2014,hur17b}.

The paper is organized as follows:
In Sect.~\ref{secth}, we present the theoretical framework for the tSZ effect, the mass accretion rate, and the galaxy cluster mass function evolution.
Then, in Sect.~\ref{secmeth}, we detail the MAR extraction methodology.
Finally, we present and discuss the result in Sects.~\ref{secres}~and~\ref{secconcl}.\\
In the following, we consider $Planck$-CMB best fitting cosmology \citep{planckpar18} for the 6 parameter concordance standard model.

\section{Theoretical framework}
\label{secth}

\subsection{Thermal Sunyaev-Zel'dovich effect}

The tSZ effect \citep{sun72} is a tailored mass proxy for galaxy clusters \citep{planckszs}.\\
This effect is a distortion of the CMB blackbody radiation through inverse Compton scattering. CMB photons receive an average energy boost when scattering off hot (a few keV) ionized electrons of the intra-cluster medium \citep[see e.g.,][for reviews]{bir99,car02}.
The intensity of the tSZ effect in a given direction on the sky is measured by the thermal Compton parameter, $y$, which is related to the electron density along the line of sight by
\begin{equation}
y (\vec{n}) = \int n_{e} \frac{k_{\rm{B}} T_{\rm{e}}}{m_{\rm{e}} c^{2} } \, \sigma_{T} \  \mr{d}s
\label{comppar}
,\end{equation}
where $\mr{d}s$ is the distance along the line of sight,  $n_{\rm{e}}$
and $T_{e}$ are respectively the electron number density and temperature. In units of CMB temperature the contribution of the tSZ effect for a given observation frequency $\nu$ is
\begin{equation}
\frac{\Delta T_{\rm{CMB}}}{T_{\rm{CMB}} }= g(\nu) \ y.
\end{equation}
where, neglecting relativistic corrections, we have the frequency factor \begin{equation}
g(\nu) = \left[ x\coth \left(\frac{x}{2}\right) - 4 \right] \quad \mr{with} \quad x=\frac{h \nu}{k_{\rm{B}} \, T_{\rm{CMB}}}
\label{szspec}
,\end{equation}
where $T_{\rm CMB}$~=~2.726$\pm$0.001~K, the tSZ effect is negative below 217~GHz and positive for higher frequencies.\\


\subsection{Mass accretion history}

To model the relation between the galaxy cluster MAR and the cosmological parameters, we used the results from \citet{cor15},
where the galaxy cluster mass accretion history is described by
\begin{align}
\label{eqmah}
M(z) &= M_0 (1+z)^{\alpha f(M_0)}e^{-f(M_0)z}, \\
\alpha &= 1.686 \times (2/\pi)^{1/2} \frac{{\rm d}D}{{\rm d}z}|_{z=0} + 1, \\
f(M_0) &= 1/\sqrt{S(M_0/q) - S(M_0)},\\
q &= 4.137 \tilde{z}_f^{-0.9476},\\
\tilde{z}_f &= -0.0064 \times ({\rm log}_{10} M_0)^2 + 0.0237 \times ({\rm log}_{10} M_0) \nonumber \\
& + 1.8837,
\end{align}
Where $M_0 = M(z=0)$ and 
\begin{align}
S(M) =\frac{1}{2\pi^2} \int P(k) W^2(k,R) k^2 {\rm d}k,
\end{align}
with $P(k)$ the linear power spectrum of the matter distribution and $W(k,R)$ the Fourier transform of a top hat window function of radius $R$ corresponding to the mass $M$.

\subsection{Galaxy cluster mass function}

Considering that the universe is homogeneous at large scales, the galaxy cluster mass function at $z=0$ should be the same at all locations in the Universe.\\
However, due to galaxy cluster mass accretion history, the number density of high mass galaxy clusters is higher at low-$z$.
The redshift evolution of the mass function is therefore a proxy for the galaxy mass accretion history.\\

\begin{figure}[!h]
\begin{center}
\includegraphics[width=0.9\linewidth]{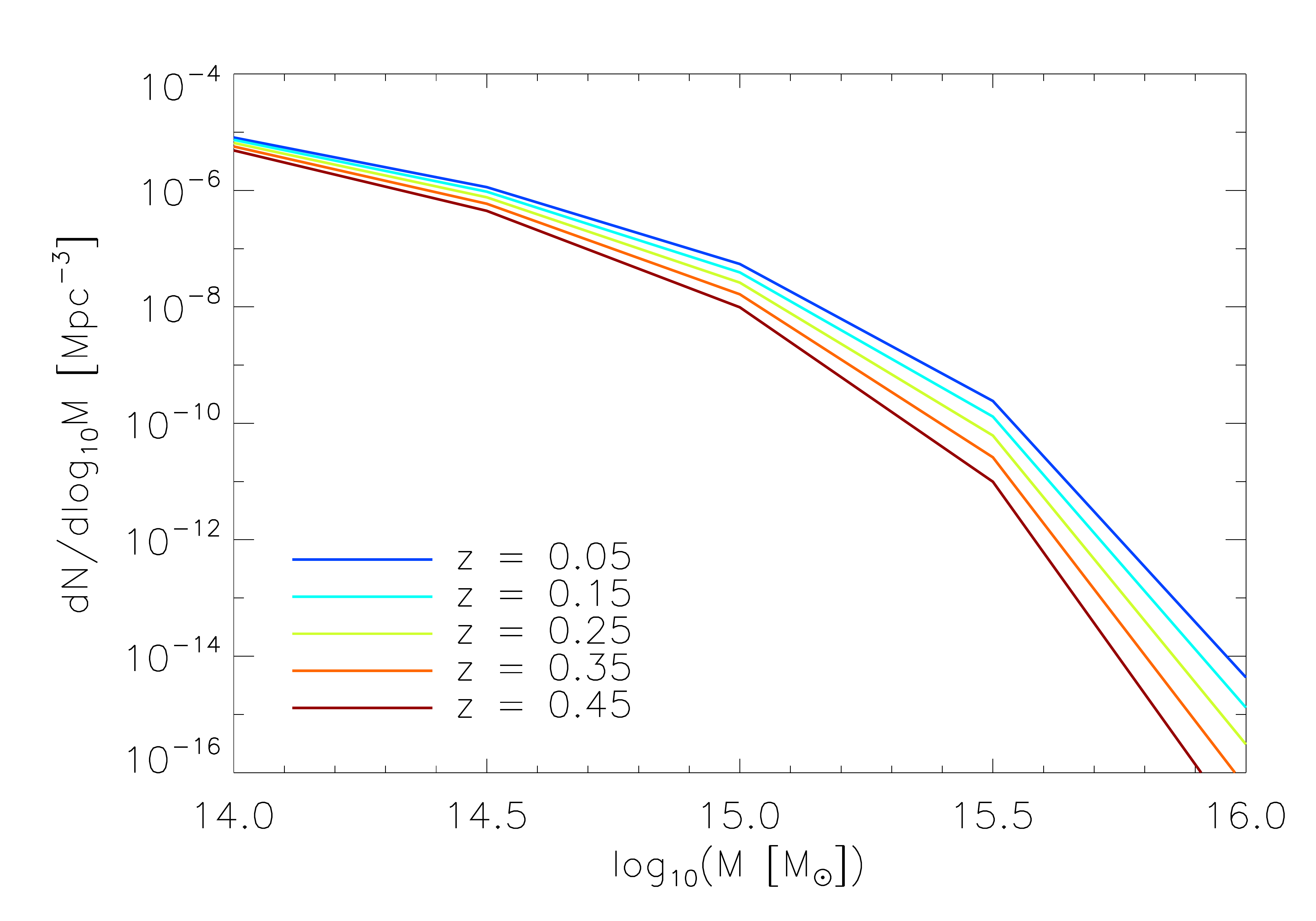}
\caption{Evolution of the galaxy cluster mass function, ${\rm d}N/{\rm d}M{\rm d}V$, for $0.0 < z < 0.5$ considering the parametric ajustement from \citet{tin08}.}
\label{figmf}
\end{center}
\end{figure}

\noindent On Fig.~\ref{figmf}, we show the mass-function evolution from $z=0.55$ to $z = 0.05$ using the mass-function parametric form from \citet{tin08}.
The mass-function is particularly steep for $M \geq 10^{15}$ M$_\odot$. Consequently, the high-end of the mass function is particularly sensitive to the mass accretion history of galaxy clusters. This mass function is presented as an illustration, in the following, we are not using any theoretical mass function.

\section{Extraction of the mass accretion from galaxy cluster number count}
\label{secmeth}

\subsection{Data}

we use the $Planck$ SZ source catalogue \citep[PSZ2
  hereafter, see ][]{psz2}. It consists of 1653 sources
detected through their tSZ effect in the $Planck$ frequency maps.
We also use the $Q_{\rm NEURAL}$ parameter \citep{nab15} to assess the quality of the tSZ sources candidate.
Follow-up results from \citet{van16} have shown the accuracy of this parameter to disentangle galaxy clusters from spurious candidates.\\

\begin{figure}[!h]
\begin{center}
\includegraphics[width=0.9\linewidth]{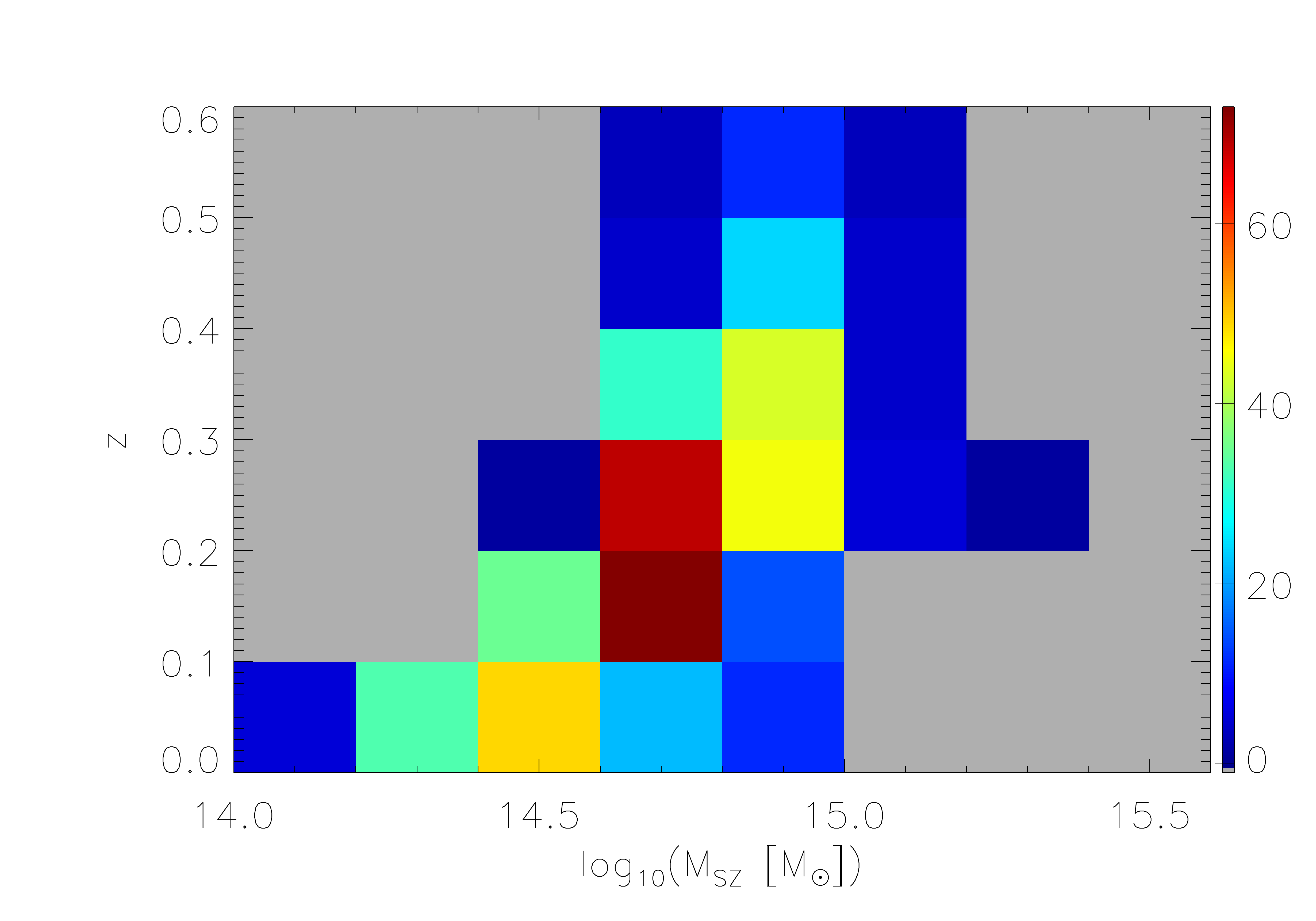}
\caption{Number of PSZ2 objects with S/N$_{\rm MMF3} > 7$ and $Q_{\rm NEURAL} > 0.5$ in for various bins in the $M$-$z$ plane.}
\label{figcov}
\end{center}
\end{figure}

\noindent On Fig.~\ref{figcov}, we present the $M$-$z$ plane coverage of the PSZ2 galaxy clusters detected with a signal to noise S/N$_{\rm MMF3} > 7$ and $Q_{\rm NEURAL} > 0.5$.
We note that most of the clusters have masses $10^{14.5} < M_{\rm SZ}< 10^{15.0}$ M$_{\odot}$ and $z < 0.5$.
The low-mass high-$z$ part of this plane is not populated due to selection effects (galaxy clusters with S/N$_{\rm MMF3} < 7$), whereas the high-mass part of the plane is not populated due to the lack of such massive objects in the universe.

\subsection{relative abundance matching}

First, we assume that the tSZ effect-mass relation logarithmic scatter is invariant with the redshift.
We also assume that the evolution of the tSZ effect-mass scaling relation is properly accounted for the derivation of tSZ masses in the PSZ2 catalogue.
Thus, for a given set of cosmological parameters ($\Omega_{\rm m}$, $\sigma_8$, $H_0$), we evolve galaxy clusters from their $M_{\rm SZ}(z)$ mass to their expected mass at $z=0$ accordingly to the mass accretion history from Eq.~\ref{eqmah}.\\
Then, we build the mass functions as a function of $M_{\rm SZ}(z=0)$, ${\rm d}N/_{\rm obs}{\rm d\, log_{10}}(M_{\rm SZ})$, for $ z \in [0.0,0.1[$, $[0.1,0.2[$, $[0.2,0.3[$, $[0.3,0.4[$, $[0.4,0.5[$, and $[0.5,0.6[$.\\
For each galaxy cluster, we use the measured apparent size, $\theta_{500}$, and measured integrated Compton parameter, $Y_{500}$, \citep{psz2} to correct each object for the completeness of the PSZ2 catalogue.\\
We also build the total mass function for $z \in [0.0,0.6[$, ${\rm d}N_{\rm tot}/{\rm d\, log_{10}}(M_{\rm SZ})$. From this mass function, we compute the expected corresponding number of object, $N$, as a function of $M_{\rm SZ}$ and $z$,
\begin{align}
N_{\rm exp}(M_{\rm SZ},z) = \frac{{\rm d}N_{\rm tot}}{{\rm d\, log_{10}}(M_{\rm SZ})}\,  {\rm d\, log_{10}}(M_{\rm SZ}) \, {\rm d }V(z) \, F_{\rm sel}(M_{\rm SZ},z),
\end{align}
where ${\rm d }V(z)$ is the comoving volume element at redshift $z$ and $F_{\rm sel}(M_{\rm SZ},z)$ is the selection function of the PSZ2 galaxy cluster catalogue.\\
For each bin in the $z$-$M$ plane we compute error bars assuming Poisson statistic.
Then for each set of cosmological parameters we compute the corresponding $\chi^2$,
\begin{align}
\chi^2 = \sum \left( \frac{N_{\rm obs}(M_{\rm SZ},z) - N_{\rm exp}(M_{\rm SZ},z)}{\sqrt{N_{\rm exp}(M_{\rm SZ},z)}} \right)^2.
\end{align}
This test is an assessment of the assumption that the mass function (expressed as a function of $M_{\rm SZ}(z=0)$) is invariant as a function of redshift. Which implies that the corresponding cosmological model account properly for the mass accretion history.\\

\begin{figure}[!h]
\begin{center}
\includegraphics[width=0.9\linewidth]{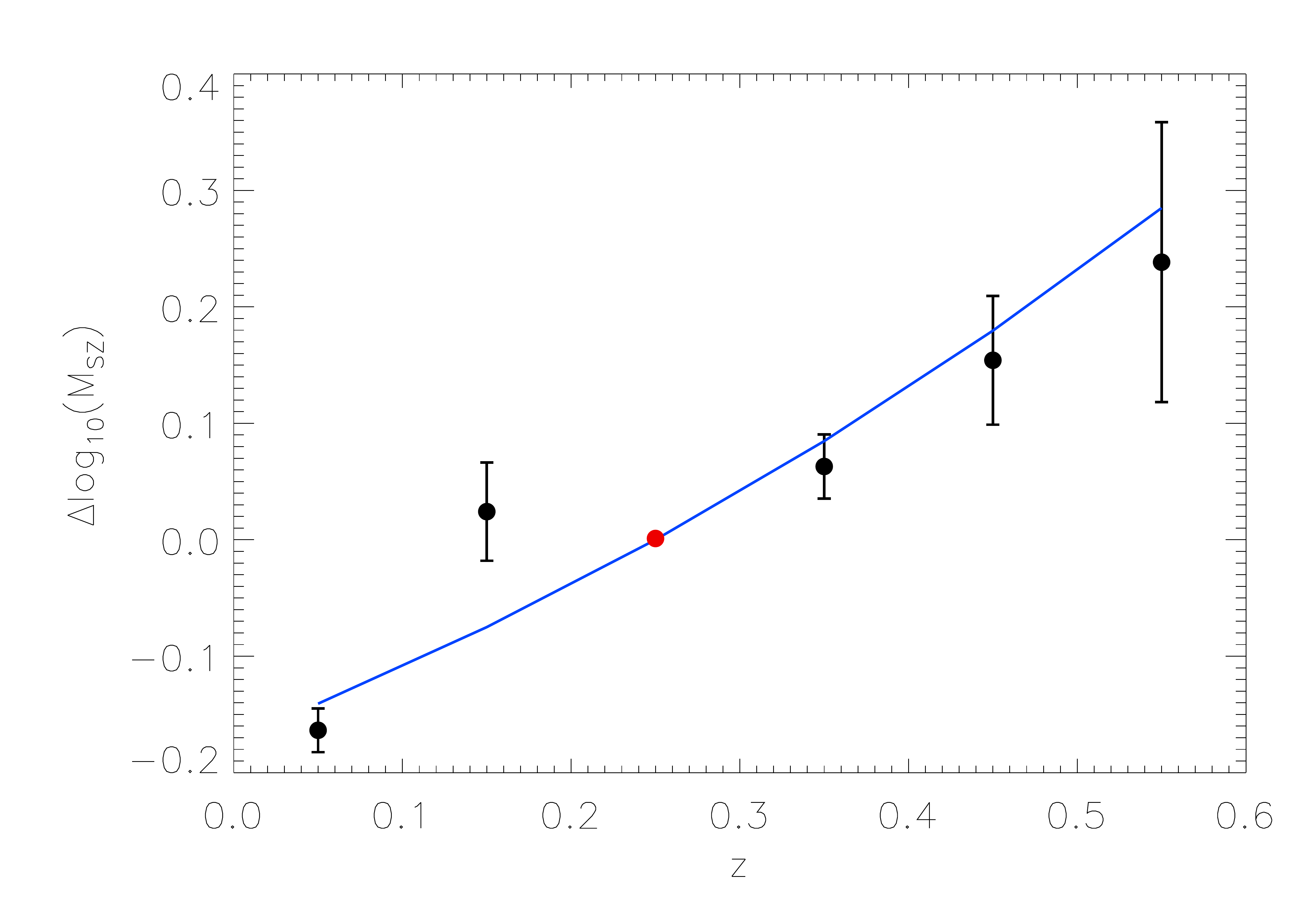}
\caption{Mean relative mass variation with respect to $M_{\rm SZ}(z=0.25)$ as a function of the redshift. The solid blue curve shows the best-fitting mass-accretion history model.}
\label{figmes}
\end{center}
\end{figure}

On Fig.~\ref{figmes}, we show the mean relative mass variation compared to $M_{\rm SZ}(z=0.25)$, 
\begin{align}
\Delta {\rm log_{10}}(M_{\rm SZ}) =  {\rm log_{10}}(M_{\rm SZ}(z = 0.25)) - {\rm log_{10}}(M_{\rm SZ}(z)).
\end{align}
The choice of normalization at $z = 0.25$ is arbitrary as we are just focussing on the mass function evolution.\\
Each data sample on Fig.~\ref{figmes} is derived by computing the mean mass variation needed to match the mass function at redshift $z$ with respect to the mass function at $z=0.25$.\\

\section{Results}
\label{secres}

For the analysis, we vary $\sigma_8$, $\Omega_{\rm m}$, and $H_0$ without priors, whereas all other parameters are marginalized as nuisance parameters. The constraints are highly degenerated for these three parameters. Consequently, we build a combined parameter of the form $\sigma_8 \Omega_{\rm m}^{\alpha} H_0^{\beta}$. We determine the $\alpha$ and $\beta$ parameters by minimizing the likelihood function dispersion. We derive $\alpha = -0.3$ and $\beta = -0.2$, whereas for the tSZ angular power spectrum this degeneracy is $\sigma_8 \Omega^{0.3}_{\rm m}$.\\

\begin{figure}[!h]
\begin{center}
\includegraphics[width=0.9\linewidth]{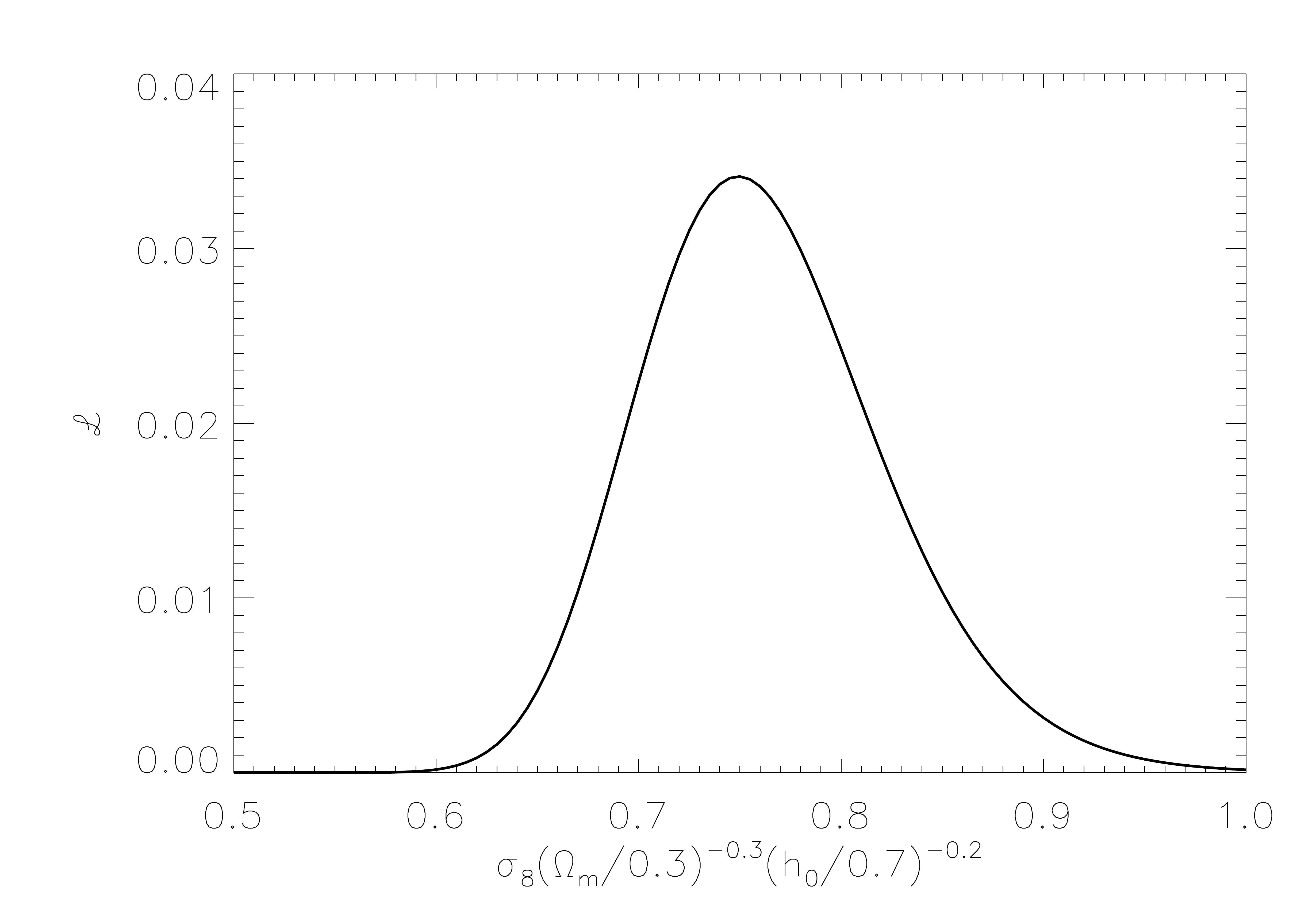}
\caption{Likelihood function for $\sigma_8 \Omega_{\rm m}^{-0.3} H_0^{-0.2}$ derived from the MAR reconstructed from galaxy cluster number count evolution.}
\label{figdet}
\end{center}
\end{figure}

On Fig.~\ref{figdet}, we present the likelihood function for the parameter $\Sigma_8 = \sigma_8 (\Omega_{\rm m}/0.30)^{-0.3} (H_0/70)^{-0.2}$, we derive $\Sigma_8 = 0.75 \pm 0.06$.\\
The global best fit is presented on Fig.~\ref{figmes} (solid blue line) and illustrates the very good agreement (within $\sim2\sigma$) among all measurements for individual redshifts.
We obtain a $\chi^2_{\rm ndf}$ of $1.6$ for 5 degrees of freedom for the best fitting model.\\
This results is consistent with tSZ galaxy cluster number counts, tSZ angular power spectrum, and tSZ bispectrum analyses \citep{hur17b}.
It is also consistent, with cosmic microwave background weak lensing, X-rays, and tSZ cross-correlations \citep{hur15a,hur15b,hur17c}.
It is also compatible with results derived from the cosmic microwave background \citep{planckpar18}.\\

\noindent Due to sMAR nearly independence with mass, the sensitivity to hydrostatic mass bias is small, a variation of 20\% for the mass bias induce only a variation of 0.02 over $\Sigma_8$, with higher value for the mass bias favoring higher values for $\Sigma_8$.
For this work, we assumed $(1-b_{\rm H}) = 0.74$ \citep{hur18}.

\section{Conclusion}
\label{secconcl}

We have presented a new approach to set cosmological constraints from galaxy cluster number count by focussing only on the evolution of the galaxy cluster mass function to extract constraints of the mass accretion history of galaxy clusters.\\
This technic probes the average mass evolution of the galaxy clusters, ${\rm d}M/{\rm d}z$, under the assumption that the universe is homogeneous at large scales.
Therefore, it does not assume any parametric form for the galaxy cluster mass-function.
It is particularly sensitive to the high-end of the galaxy cluster mass function and is therefore not strongly sensitive to the selection function of the galaxy cluster sample\footnote{Considering that the cluster sample is complete for high-mass objects.}.\\

\noindent The constraints derived from this approach, on $\sigma_8 \Omega_{\rm m}^{-0.3} H_0^{-0.2}$, are following significantly different degeneracies than typical number count analyses or angular power spectrum analyses, $\sigma_8 \Omega_{\rm m}^{0.3}$.\\
Additionally, due to the very small dependance of the sMAR with the galaxy cluster mass, this probe allows to be nearly insensitive to the mass-observable relation calibration, which is presently one of the main limitations in the cosmological exploitation of galaxy clusters \citep[see e.g.,][]{hur15b}.
However, this probe remain sensitive to the slope of the mass-observable relation.\\

\noindent Beyond the constraints on cosmological parameter, this tomographic approach of the galaxy cluster mass function allows to test the consistency between the normalization of galaxy cluster number count (which dominates usual constraints from number count) and its evolution.
We have shown on Fig.~\ref{figmes} that all redshift bins presents a coherent evolution for the mass function. We also note that the best-fitting cosmological parameters also favors low values for $\Sigma_8$. Which implies, considering the degeneracy relation, a low-value for $\sigma_8$, or high values for $\Omega_{\rm m}$ and/or $H_0$.\\

\noindent We derived $\sigma_8 (\Omega_{\rm m}/0.30)^{-0.3} (H_0/70)^{-0.2} = 0.75 \pm 0.06$, which is compatible with other galaxy cluster related analyses and cosmic microwave background analyses.\\
As this method relies on the assumption that the mass-observable relation evolution is properly accounted in the mass derivation, this results also imply that the tSZ mass-observable relation is not significantly deviating from its expected redshift evolution at $z < 0.6$.\\

\noindent recently, the MAR, ${\rm d}M/{\rm d}t$, has also been measured using galaxy cluster virial shock of individual cluster such as A2319 \citep{hur19}. Future microwave and sub-millimeter data will allow such measurements for large sample of galaxy clusters.\\
Combining mass-function evolution with individual mass accretions rates will provide a new way of probing the Hubble parameter (time-redshift relation) with galaxy clusters.

\section*{Acknowledgment}
\thanks{G.H. acknowledge support from Spanish Ministerio de Econom\'ia and Competitividad (MINECO) through grant number AYA2015-66211-C2-2. U.K. acknowledges support by the Israel Science Foundation (ISF grant No. 1769/15)}

\bibliographystyle{aa}
\bibliography{cosmo_mar}

\end{document}